\documentclass[twocolumn,amsmath,amssymb]{revtex4}
\usepackage[]{graphicx}
\usepackage{dcolumn}
\usepackage{bm,multirow}

\begin{document}

\title{All-optical conditional logic with a nonlinear photonic crystal nanocavity}

\author{Murray W. McCutcheon}
\altaffiliation{experiments were conducted at the University of British Columbia}
\affiliation{School of Engineering and Applied Sciences, Harvard University, Cambridge, MA 02138}
\author{Georg W. Rieger, and Jeff F. Young}
\affiliation{Department of Physics and Astronomy, University of British Columbia,
Vancouver, Canada, V6T 1Z1}
\author{Dan Dalacu}
\affiliation{Institute for Microstructural Sciences, National Research Council, Ottawa, Canada, K1A OR6}
\author{Philip J. Poole}
\affiliation{Institute for Microstructural Sciences, National Research Council, Ottawa, Canada, K1A OR6}
\author{Robin L. Williams} 
\altaffiliation[also at ]{Department of Physics, University of Ottawa, Ottawa, Canada, K1N 6N5.}
\affiliation{Institute for Microstructural Sciences, National Research Council, Ottawa, Canada, K1A OR6}

\date{\today}

\begin{abstract}  We demonstrate tunable frequency-converted light mediated by a $\chi^{(2)}$
nonlinear photonic crystal nanocavity.  The wavelength-scale InP-based cavity supports 
two closely-spaced localized modes near 1550 nm which are resonantly excited by a 130 fs 
laser pulse.  The cavity is simultaneously irradiated with a non-resonant probe beam, giving 
rise to rich second-order scattering spectra reflecting nonlinear mixing of the different
resonant and non-resonant components.  In particular, we highlight the radiation at the sum 
frequencies of the probe beam and the respective cavity modes.  This would be a useful,
minimally-invasive monitor of the joint occupancy state of multiple cavities in an 
integrated optical circuit.



\end{abstract}

\maketitle

To realize all-optical logic on an integrated semiconductor chip, it is necessary to develop
a toolkit of logical operations (gates) which can manipulate optical signals.  These operations
serve to generate, condition, and detect optical signals, and are most naturally applied
in cavities, where photons can be localized for many optical cycles.
Recent work has demonstrated that photonic crystal nanocavities can be fabricated with 
$Q$ factors greater than $10^6$, and are thus capable of storing photons for more than 1 
nanosecond~\cite{Tanabe_07, Takahashi_07} and realizing one-bit delays in bit streams
of data~\cite{Notomi_NPh08}. Cavities can also serve as channel drop filters to 
transfer a signal from one waveguide to another~\cite{FanPRL98}.  

In both classical and quantum optical protocols, the ability to perform 
conditional logic requires nonlinear functionality~\cite{Soljacic04, Turchette}.  
Recent advances on this front include all-optical switching via bistability and
free-carrier tuning~\cite{Cowan03, Tanabe_05,Barclay05, Fushman07, Almeida}, 
dynamic $Q$ factor control for the
release of photons on demand~\cite{Tanaka_07}, and adiabatic frequency conversion through 
the perturbation of cavity modes~\cite{McCutcheon_OE, Tanabe_09, Preble07, NotomiPRL}.  
The latter approach is effective for generating small frequency shifts in an optical 
mode, but for larger shifts, the nonlinearity of the material can be exploited to 
generate new harmonics.  Theoretical work has shown that, in principle, 
loss-less harmonic generation is feasible in doubly-resonant $\chi^{(2)}$ and $\chi^{(3)}$ 
cavities~\cite{Rodriguez07,Burgess09}.  Experimentally, we have shown in our previous 
work that second-harmonic generation and sum-frequency generation are possible at 
microwatt-scale powers by leveraging the high $Q$ factor and small mode volume of 
a photonic crystal nanocavity~\cite{McCutcheon_PRB}.   

To perform conditional logic, it is important to be able to
monitor mode occupation in a cavity.  In this letter, we use nonlinear mode mixing in
a photonic crystal nanocavity to demonstrate a method to monitor
the populations of modes in {\em different} cavities in a weakly perturbative fashion.  
The experiment involves conditionally 
generating a signal dependent both on the occupation of one microcavity mode (or many)
and the overlapping presence of a transient (non-resonant) signal in the cavity. In addition
to serving as a joint-state monitor, this technique could be used to generate new signals
at a frequency of choice.


\begin{figure}[htbp]
\centering
\includegraphics[width=6cm]{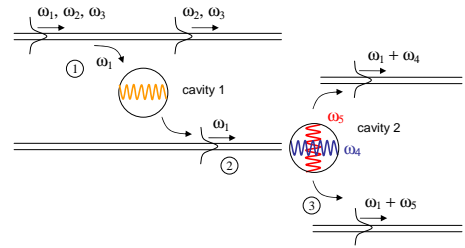}
\caption[Schematic optical circuit for the joint state nonlinear monitor]
{\label{fig:tune}Schematic optical circuit 
for the joint state nonlinear monitor.  (1) Cavity 1 drops channel $\omega_1$ from 
a multi-mode waveguide. (2) $\omega_1$ is
coupled into cavity 2 (non-resonantly).  Cavity 2 supports two resonant modes, 
$\omega_4$ and $\omega_5$, and when $\omega_1$ arrives, (3) new frequencies are
conditionally generated at $\omega_1+\omega_4$ and/or $\omega_1+\omega_5$, depending 
on the population of cavity 2.
}
\end{figure}

Our experiment is motivated by the schematic optical circuit of Fig.~\ref{fig:tune}.
The figure shows two cavities, {\bf 1} and {\bf 2}, with {\bf 1} acting as a 
single mode channel drop filter~\cite{FanPRL98}, 
and {\bf 2} acting to store photons at frequencies $\omega_4$ and
$\omega_5$.  Signal $\omega_1$ is dropped from a multi-mode channel through 
the microcavity filter and into the middle waveguide.  When signal $\omega_1$
reaches cavity {\bf 2}, it can interact with modes $\omega_4$ and $\omega_5$ via
the material second-order nonlinearity of the cavity.  
New signals are generated at $\omega_1 + \omega_4$ and/or
$\omega_1 + \omega_5$ that are conditional on the combined occupation state
of cavities {\bf 1} and {\bf 2}.  Our experiment illustrates the nonlinear physics 
involved in generating these conditional signals.

The slab which hosts the two-mode PhC nanocavity is a 230 nm thick [001]-oriented InP 
free-standing membrane mounted on a glass substrate.  It is excited by a train of 
$\sim$ 130 fs pulses from a 80 MHz optical parametric oscillator (OPO) (Spectra Physics) 
with a bandwidth of 100 cm$^{-1}$ and an average power of 60 $\mu$W that is
focussed through a 100$\times$ microscope objective.  Using our well-established 
technique~\cite{McCutcheon_APL}, the resonantly scattered light is collected 
in reflection, and detected in the cross-polarization with respect to the incident beam
using a Bomem Fourier transform infrared spectrometer.  The second-order radiation 
is collected in transmission using a 40$\times$ (NA=0.65) 
microscope objective, and detected using a grating spectrometer and a liquid-nitrogen 
cooled CCD detector.

\begin{figure}[htbp]
\centering
\includegraphics[width=6cm]{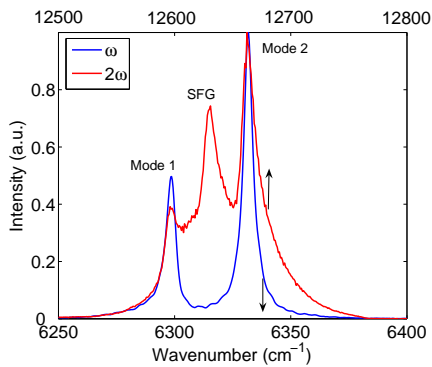}
\caption{Linear (blue) and second-order (red) scattering spectra from a 2-mode cavity.
\label{fig:2mode}}
\end{figure}

We first consider the {\em resonant} linear and nonlinear response of the cavity (taken to
be cavity {\bf 2} in Fig.~\ref{fig:tune}).   The whole spectrum is excited simultaneously 
due to the wide bandwidth of the OPO, and the modes show up as resonant peaks.
The linear spectrum is shown in blue in Fig.~\ref{fig:2mode}(b), clearly 
revealing the two modes 
of interest.  These modes aren't part of the usual in-bandgap spectrum of the L3-cavity, 
but rather are quasi-localized modes in the dielectric band continuum~\cite{McCutcheon_PRB}.  
They suit our purposes well because of their close spectral spacing (which allows 
simultaneous excitation by a single short pulse) and relatively high 
$Q$ factors.  The nonlinear spectrum, superimposed on the same plot for comparison 
purposes, is shown in red.  The lowest and 
highest energy peaks are at exactly twice the frequencies of the microcavity modes evident 
in the linear spectra, and the central sharp feature, at 2 $\times$ 6315 cm$^{-1}$, is 
at precisely their sum frequency.  The extra peak at the sum frequency therefore 
corresponds to the second-order nonlinear interaction of fields resonant in the
two modes, and suggests an application.  In an integrated optical circuit, detection of 
this second-order radiation using narrow band filters would provide a weak, 
non-destructive means of monitoring the joint occupation state of the 
microcavity, providing a logical AND gate operation.  Although this functionality
is demonstrated here in the case where a single short pulse is used to simultaneously
populate the two modes, the sum-frequency signature would also occur if the
modes were populated by independent sources.

\begin{figure}[ht]
\centering
 \includegraphics[width=5cm]{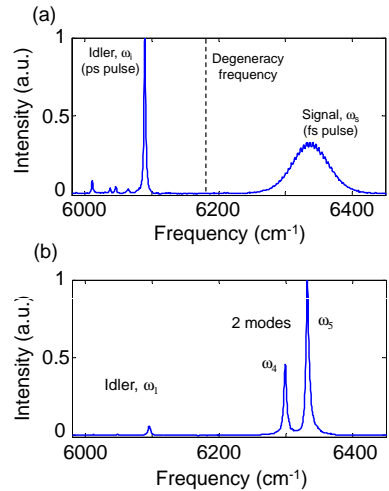}
\caption{\label{fig:sigidler}(a)  Spectrum of the OPO laser beam, showing the
femtosecond (fs) pulse at 6335 cm$^{-1}$ and the prominent picosecond (ps) pulse
at the frequency of the idler.  A number of smaller features are also visible
near the main idler peak.  (b) Resonant scattering spectrum from the microcavity
studied in Fig.~\ref{fig:idler}, using the source as in (a), but slightly detuned.
}
\end{figure}

To demonstrate the conditional mode monitoring operation, we simultaneously irradiate
the cavity with short, resonant pulses (as above) and a second {\em non-resonant} excitation beam
consisting of longer, picosecond pulses tuned far off resonance with the cavity modes.
This two-color source is readily available from the  unfiltered ``signal'' beam output of the 
OPO when it is tuned near the degeneracy point (where both signal and idler frequencies are 
close to half the pump frequency).   An example of this unfiltered spectrum when the signal 
is tuned to 6335 cm$^{-1}$ is shown in Figure~\ref{fig:sigidler}(a). 
The short OPO signal pulses are accompanied by relatively long (a few ps) OPO idler pulses.
The idler pulses in this spectrum
are dominated by a single spectral feature at $\omega_i = 6090$ cm$^{-1}$, 
but there are several smaller intensity peaks at lower frequency which are also revealed 
in the second-order nonlinear processes discussed below.  The resonant scattering 
spectrum from the two-mode microcavity studied in Fig.~\ref{fig:2mode}, when excited by
the source of Fig.~\ref{fig:sigidler}(a), is shown
in Fig.~\ref{fig:sigidler}(b).  The resonant enhancement of the mode features is
evident in comparison to the non-resonantly scattered idler.

\begin{figure}[htb]
\centering
\includegraphics[width=9cm]{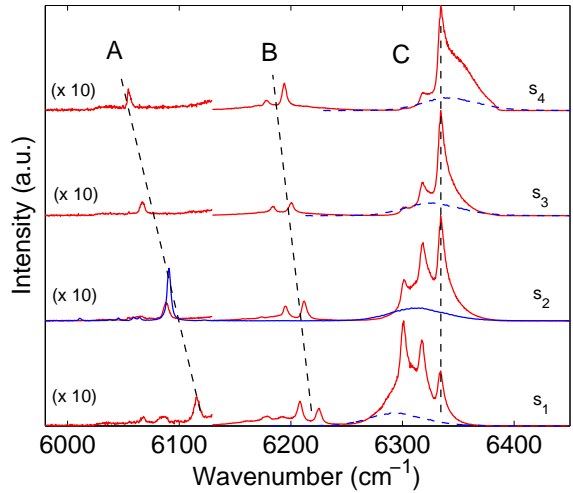}
\caption{\label{fig:idler}Spectra given by the interaction of a non-resonant,
narrow linewidth idler pulse; a broad, resonant laser pulse; and a two-mode PPC microcavity. 
The four second-order spectra (red) are plotted at half the measured energy.  
The signal pulse is tuned to higher energy for each spectrum s$_1$-s$_4$, and so the 
idler tunes to lower energy.  The amplitude of the low energy region has been 
multiplied by 10 for clarity.  The solid blue curve shows the laser spectrum 
scattered from a non-textured region of the sample, as in Fig.~\ref{fig:sigidler}(a).  
The dashed blue curves schematically show the signal pulse spectrum tuning to higher
energies.}
\end{figure}

We now use this spectrum to explore the scheme shown in Fig.~\ref{fig:tune}.
The off-resonant ps pulse from the idler plays the role of $\omega_1$, while the two 
modes are associated with $\omega_4$ and $\omega_5$.
Figure~\ref{fig:idler} shows a series of four second-order spectra from the microcavity 
supporting the quasi-localized modes discussed in Fig.~\ref{fig:2mode}(b) irradiated
by the two-color source.  The OPO is tuned to four different values of the center
wavelength to acquire the spectra, which show three principal groups of features that 
are marked by lines 
A, B, and C to guide the eye. Feature A and the broad background in group C (the fit of 
which is plotted separately as a dashed blue line), are due to non-resonant 
second-harmonic generation of the ps and fs features, respectively, in the excitation 
spectra.  

The three sharp features in group 
C that do not shift, and the two sharp features in group B that shift at half the rate 
of the excitation beam(s), are specific to the microcavity modes.  They all arise from 
second-order processes involving the fields ``trapped'' in at least one of the modes.  
The three (fixed) peaks in group C correspond to the mode SHG 
($2\omega_4$, $2\omega_5$) and SFG ($\omega_4 + \omega_5$) features, 
as in Figure~\ref{fig:2mode}.  The features in group B are then easily
understood to result from the second-order radiation from the two field distributions 
trapped in the microcavity modes respectively interacting with the ps pulses that 
irradiate the cavity during the ring-down to produce peaks at $\omega_1 + \omega_4$ and
$\omega_1 + \omega_5$.  To understand the difference
in the shift rate between features A and B, consider a ps pulse at $\omega_A$ interacting
with a microcavity mode at $\omega_C$.  When the ps pulse is tuned from $\omega_A$
to $\omega_A - \Delta\omega$, the second-order feature A shifts from 2$\omega_A$ to
$2(\omega_A-\Delta\omega)$, which is a shift of $-2\Delta\omega$, whereas feature B shifts
from $\omega_A+\omega_C$ to $\omega_A - \Delta\omega +\omega_C$, a shift of just
$-\Delta\omega$.  

The lifetime of each cavity mode is about 1 ps, and so the pulsed excitation probes
the impulse response of the cavity (i.e. there is minimal local field enhancement).
The coupling efficiency into the cavity of the 60 $\mu$W average 
power resonant excitation beam is $\sim 2\%$~\cite{Banaee07}; therefore, accounting 
for the 80 MHz repetition rate of the OPO, approximately $10^5$ photons are resonantly 
coupled into the cavity 
with each pulse.  It is useful to consider what equivalent, critically-coupled cw-excitation
or quasi-cw-excitation would create mode populations sufficient to yield an equivalent 
nonlinear response. In the cw limit, the local field intensity 
in the cavity is enhanced by a factor of $Q$.  Assuming the cw-beam is coupled into the
cavity from a 600 nm $\times$ 200 nm single-mode ridge waveguide, the 
same nonlinear response could be achieved by a cw power of about 20 $\mu$W. 

The processes illustrated here demonstrate that the fields stored in
microcavity modes can be nonlinearly mixed with non-resonant signals to produce 
sum-frequency radiation.  If the non-resonant fields were generated 
from light previously stored in a different microcavity, as in the schematic of
Fig.~\ref{fig:tune}, this principle could be used to conditionally
generate information at new frequencies which depend on the joint occupation
state of two different microcavities.
\vspace{-10pt}
\section*{Acknowledgements}
The authors  wish to  acknowledge the financial support of the Natural Sciences and
Engineering Research Council of Canada, the Canadian Institute for Advanced Research,
the Canadian Foundation for Innovation, and the Canadian Institute for Photonic Innovations.


\end{document}